**How far are Extraterrestrial Life and Intelligence after Kepler ?**

Amri Wandel, the Hebrew University of Jerusalem
E-mail: amri@huji.ac.il

**Abstract**

The Kepler mission has shown that a significant fraction of all stars may have an Earth-size habitable planet. A dramatic support was the recent detection of Proxima Centauri b. Using a Drake-equation like formalism I derive an equation for the abundance of biotic planets as a function of the relatively modest uncertainty in the astronomical data and of the (yet unknown) probability for the evolution of biotic life, $F_b$. I suggest that $F_b$ may be estimated by future spectral observations of exoplanet biomarkers. It follows that if $F_b$ is not very small, then a biotic planet may be expected within about 10 light years from Earth. Extending this analyses to advanced life, I derive expressions for the distance to putative civilizations in terms of two additional Drake parameters - the probability for evolution of a civilization, $F_c$, and its average longevity. Assuming "optimistic" values for the Drake parameters, $(F_b \sim F_c \sim 1)$, and a broadcasting duration of a few thousand years, the likely distance to the nearest civilizations detectable by SETI is of the order of a few thousand light years. Finally I calculate the distance and probability of detecting intelligent signals with present and future radio telescopes such as Arecibo and SKA and how it could constrain the Drake parameters.

**1. Introduction**

An important yet until recently poorly known factor, required for estimating the abundance of extraterrestrial life is the fraction of stars with planets, in particular Earth-like planets within the Habitable Zone. In the past decade planets have been discovered around hundreds of nearby stars, yet most of them were Jupiter-like gas giants, and too close to their host star to permit liquid water on their surface (e.g. Fridlund *et al.* 2010). In the last four years, the Kepler mission yielded over 5000 exoplanet candidates, most of them with sizes smaller than Neptune and down to Earth-sized planets (~1-2 Earth radii)**.** Such small planets have been shown to constitute the majority of exoplanets (Buchhave et al., 2012; Batalha et al., 2013; Dressing and Charbonneau, 2015). Further work has demonstrated that such planets are often found within the Habitable Zone of their host star. Recent analyses of the Kepler data showed (Petigura *et al.,* 2013) that about 20% of all solar-type stars have small, approximately Earth-sized planets orbiting within their Habitable Zone. Observational uncertainties and false-positive detections (Foreman-Mackey, Hogg and Morton, 2014; Farr, Mandel and Stroud, 2014) may significantly reduce this figure (down to 2-4%, however with a large uncertainty), yet it still implies a significant fraction and a huge number of stars with Earth-size habitable planets. Similar results have been obtained by different methods. The HARPS team (using the Doppler method) estimated that more than 50% of solar-type stars harbor at least one planet, with the mass distribution increasing toward the lower mass end (<15 Earth masses) (Mayor *et al*., 2011). HARPS also detected Super Earths in the Habitable Zone (Lo Curto *et al.*, 2013). These findings demonstrate that "Earthlike" planets (in the sense of Earth-size planets in the Habitable Zone) are probably quite common, enhancing the probability of finding planets with conditions appropriate for the evolution of biological life as we know it.

Another conceptual break-through has been the recognition that life need not be limited to planets orbiting solar-type stars, which are only a small fraction of all stars in the Galaxy. Especially Red Dwarf (RD) or M-type stars (the lowest-mass stars, 0.08-0.6 $M_{Sun}$, less luminous and cooler than the Sun), are of great interest, as RDs are the type most common in the Galaxy, constituting about 75% of the stars in our neighborhood. In this review we focus our discussion on planets of RD stars (RDPs).

The Kepler results show that a significant fraction of all RDs may have Earth-sized planets (Dressing and Charbonneau, 2013). A recent estimate of the occurrence of terrestrial sized planets (0.5-1.4 Earth radii) in the Habitable Zones around Kepler RDs gives that 20-40% are expected to harbor an Earth-sized planet in their Habitable Zone (Kopparapu, 2013). If correct, this would mean that about every third star in the Galaxy has a terrestrial planet within its Habitable Zone. The recent detection of Proxima Centauri b [Anglada-Escudé, *et al.* 2016] supports this result.

As the origin of life on Earth is very poorly understood, some works argued that Earthlike compex life is extremely scarce (the "Rare Earth Hypothesis"; Ward and Brownlee 2000). However, 16 years later, several of the assumptions in this work were found or shown to be inaccurate: terrestrial planets in the habitable zone, even of G-stars, were found to be frequent, M stars were shown to be less hostile to host planets with conditions life than previously thought (e.g. Tarter et al. 2007; Seager and Demming 2010; Gale and Wandel 2016), and planets in the Habitable Zone of RDs were argued to be able to support liquid water and photosynthesis for a wide range of atmospheric and other properties (Wandel 2016). If RD planets are suitable for life this would imply that about half of the stars in the Galaxy could harbor a biotic planet and correspondingly enhance the potential abundance of civilizations. These findings demonstrate that Earth-like planets are probably quite common, enhancing the probability to find planets with conditions appropriate for the evolution of biological life as we know it. Based on these results we can better estimate the abundance of life in our stellar neighborhood [Wandel 2011; 2013, 2015].

In both approaches to the search for life outside of the Solar system, (i) looking for biotic signatures (biomarkers) in the spectra of Earth-like extra solar planets, and (ii) searching for intelligent electromagnetic signals (SETI), the success probability depends critically on the distance to the nearest candidates. Complex life and in particular intelligent life is presumably much scarcer than simple, mono-cellular life (e.g. Mayr 2004), and hence the distances to the nearest broadcasting civilizations may be very large, possibly beyond the detection range of present, and perhaps even future radio telescopes. In order to assess the feasibility of detecting biotic planets and intelligent extraterrestrial signals it is essential to estimate the distances to the potential targets. This work applies the recent results from the Kepler mission in order to derive useful expressions and figures for the distance to biotic exoplanets (section 2) and to putative civilizations (section 3). It is further suggested that the probability for the evolution of biotic life may be estimated by future spectral observations of exoplanet biomarkers. In a similar way it is argued (section 4) that the abundance of putative technological civilizations may be constrained by future planned radio telescopes arrays, such as SKA.

**2.1 The abundance of biotic planets**

How common are worlds harboring life? The recent findings of Kepler indicate that Earth-sized planets may be found around almost every star. However, assuming that life may develop only on Earth-like planets orbiting Sun-like stars (an assumption likely to be too conservative, as alien life may develop in environments very different from Earth's biosphere), the number of candidates may be reduced to about 10% of all stars.
As is well known, the Drake equation (eq. 5 below) estimates the number civilizations in our the Galaxy"). By analogy, the number of biotic planets in the Galaxy, $N_b$, may be assessed by a "biotic Drake equation"

$$N_b = R^* F_s F_p F_e n_{hz} F_b L_b. \qquad [1]$$

The first five terms are astronomical factors and the last two may be called "biotic parameters". The astronomical factors include the rate of star birth in the Galaxy, $R^*$, the fraction of stars suitable for evolution of life, $F_s$, and three "planetary" factors: the fraction of stars that have planets, $F_p$, the fraction of Earth-sized planets, $F_e$, and the number of such planets within the Habitable Zone, $n_{hz}$. These five

"astronomical factors" can be combined into a single parameter $R_b$, the rate at which stars suitable for the evolution of biotic life are formed in the Galaxy,

$$R_b = R^* F_s F_p F_e n_{hz}. \qquad [2]$$

The average star birth rate in the Galaxy is $R^* \sim 10$ yr$^{-1}$ [e.g. Carroll and Ostile, 2007]. If evolution of life is assumed to be limited to stars similar to our Sun, then $F_s \sim 0.1$. However, this assumption is probably a too restrictive hypothesis. The Kepler data show that Earth size planets are frequent within the Habitable Zone of lower Main Sequence small stars [e.g. Dresing and Charbonneau, 2013], which are the majority of all stars. If life can evolve on planets of red dwarfs [Guinan and Engle, 2013, Scalo, et al., 2007] then $F_s \sim 1$ (since 75% of all stars are red dwarfs).

Recent exoplanet findings, in particularly those by the Kepler mission, suggest that probably most stars have planetary systems, hence $F_p \sim 1$. Analyses of the Kepler results shows that 7-15% of the Sun-like stars have an Earth-sized planet within their habitable zone [Petigura et al., 2014], which gives $F_e n_{hz} \sim 0.1$. If biotic life is not restricted to Earth-like planets and to the Habitable Zone (e.g. as in the case of Jupiter's moon Europa) then $F_e n_{hz}$ may be even bigger, up to order unity. Combining all these factors gives for the product of the astronomical parameters in equation [1] a probable range of $0.1 < R_b < 10$ yr$^{-1}$.

The "biotic parameters" in eq. [1] are $F_b$, the probability of the appearance of biotic life within a few billion years on a planet with suitable conditions, and $L_b$, the average longevity of biotic life in units of Gyr ($10^9$ years). Equations [1] and [2] give

$$N_b \sim 10^{11} (R_b/R^*) F_b (L_b/10), \qquad [3a]$$

Similarly, the space density of biotic planets, $n_b$, may be written as

$$n_b \sim n^* (R_b/R^*) F_b (L_b/10), \qquad [3b]$$

where $n^* \sim 0.01$ ly$^{-3}$ is the average space density of stars in the Galaxy. Considering the history of life on Earth, $L_b$ is likely to be at least a few billion years, so for stars similar to the Sun (or smaller) $L_b \sim 10$ ($10^{10}$ yr). Substituting the values of $L_b$ and $R^*$ equations [3a,b] become

$$N_b \sim 10^{10} R_b F_b, \qquad [4a]$$

$$n_b \sim 0.001 R_b F_b \text{ ly}^{-3}. \qquad [4b]$$

Since the average distance between biotic worlds is $d_b \sim n_b^{-1/3}$, eq. [4b] gives

$$d_b \sim 10 (R_b F_b)^{-1/3} \text{ ly}. \qquad [4c]$$

Eq. [4c] gives the probable distance to our nearest biotic neighbor, plotted in Fig. 1, as a function of the biotic factor $F_b$.

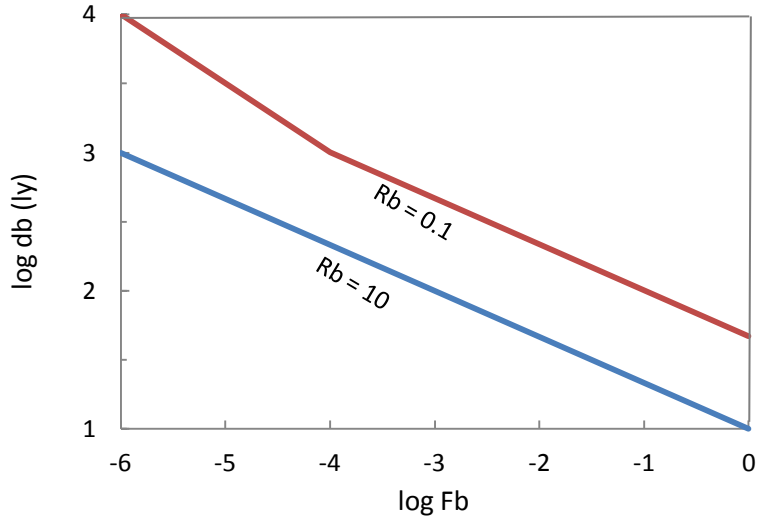

**Figure 1.** The probable distance to our nearest biotic neighbor, $d_b$ vs. the biotic factor $F_b$, for two values of the parameter $R_b$. For the break in the upper curve see the discussion following eq. [7a].

**2.2 Life on planets of Red Dwarfs**

If only planets around solar-type stars are considered, then the fraction of stars suitable for evolution of life is the well known fraction of solar-type stars in the Galaxy, $F_s \sim 0.1$ (~8% of all Main Sequence stars are G-type). According to recent analyses of the Kepler data only 2-4% of all solar-type stars have small, Earth-sized planets orbiting within their Habitable Zone (Foreman-Mackey, Hogg and Morton, 2014; Farr, Mandel and Stroud, 2014), hence for solar-type stars $F_s F_e n_{hz} \sim 0.002\text{-}0.004$. If in addition one excludes planets in multiple stellar systems (some 80% of all stars), this figure could be smaller by a factor of five, which gives the most severe constraint on the rate at which habitable planets are formed in the Galaxy, $R_b \sim 10 \times 0.2 \times 0.002 = 0.004$ per year.

However, if in addition RD stars (single or in multiple systems) may host habitable planets (e.g. Gale and Wandel 2015), then $F_s \sim 0.9$. According to the analyses of Dressing (2013) and Kopparapu (2013) up to 40% of the RD stars have habitable planets, which leads to much larger figures, $F_e n_{hz} \sim 0.4$ and $R_b \sim 4$ planets per year.

Depending on the assumptions as to which stellar type is suitable for life, the rate $R_b$ could thus vary by a factor of 1000, in the range, ~0.004 - 4. Using Eq. 3 this would yield a factor of 10 in the estimate of the distance to our nearest biotic neighbor planets. This is shown in Fig. 2, which shows the distance given by Eq. 2 as a function of the biotic factor $F_b$, for three choices of host-stars and planet-statistics, as discussed above: (a) solar type, single stars with a low fraction of Earth-sized Habitable Zone planets, ($R_b=0.004$), (b) solar type, single or multiple stars with a high fraction of Earth-sized Habitable Zone planets, ($R_b=0.1$) and (c) single or multiple Red Dwarfs ($R_b=4$). It is easy to see that assuming life can evolve on RDPs significantly enhances the abundance of potential biotic planets. For example, if $F_b = 0.1$ the estimate of the distance to our nearest biotic neighbor would vary between 140 light years in the conservative single solar-type star case and fourteen light years in the multiple RD case.

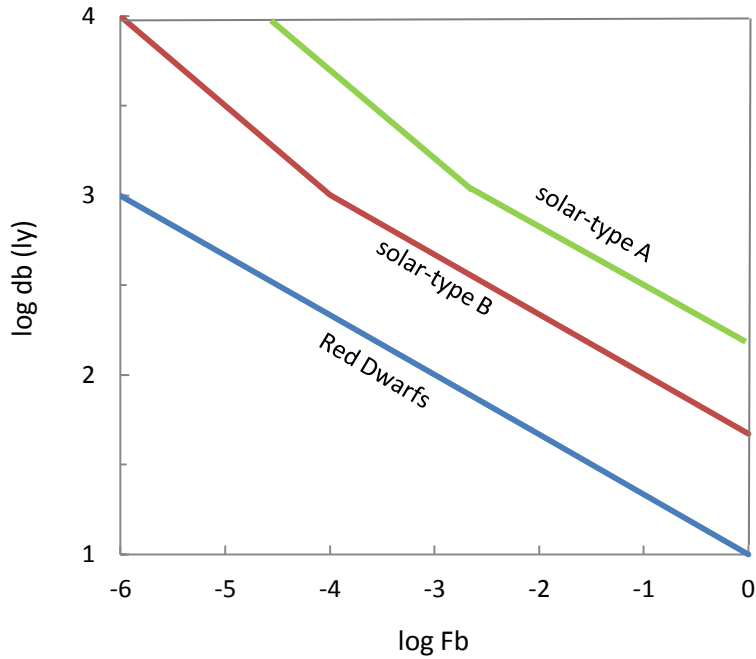

**Figure 2.** The distance to our nearest biotic neighbors as a function of the biotic factor $F_b$, for several choices of target-stars and statistics. The curve labeled "solar-type A" is for the most conservative case – only single, solar type host stars, assuming that 2% of them have Earth-sized planets in the Habitable Zone. "Solar-type B" is an intermediate case – solar type host stars in single or multiple stellar systems, assuming that 20% of them have Earth-sized planets in the Habitable Zone. The lower curve depicts single or multiple RD host stars assuming that 40% of them have Earth-sized planets in the Habitable Zone.

### 2.3. Estimating the Biotic Parameter

Given the recent progress in understanding the demographics of planets, in particular Earth-like planets (the "astronomical factor"), the biotic parameter, namely the probability of the evolution of biotic life, remains the major missing factor for estimating the distance to our neighbor biotic planets.

Simple life has appeared on Earth 3.5-3.8 Gyr ago - less than 1 Gyr after Earth's formation and merely ~0.1 Gyr after the establishment of appropriate environmental conditions (the end of the heavy bombardment), a time very short compared to the ages of Earth and the stars. *If* Earth is a typical case (the mediocrity principle), it is plausible that $F_b$ is close to unity. On the other hand, as the origin of life on Earth is very poorly understood, some experts in that area assign an extremely low probability to a similar scenario happening elsewhere (the "Rare Earth Hypothesis"; Ward and Brownlee 2000). According to this approach $F_b$ may be extremely small. However, as discussed in the section 1, several of the assumptions in the Rare Earth argumentation were found or shown to be inaccurate, in a way that suggests biotic life might be frequent.

Bayesian analysis demonstrates that as long as Earth remains the only known planet with biotic life, any value could be assigned to $F_b$ [Turner, 2012]. Hence the abundance of extraterrestrial biotic life remains an open question, depending on the (yet unknown) probability for biotic life to evolve on planets with suitable conditions within a given cosmic time. However, a breakthrough in this area may be reached in the near future, if spectral characteristics such as oxygen and other biomarker gases are detected in the atmospheres of Earth-like exoplanets, by studying transits[1] [Rauer et al. 2011; Palle et al. 2011; Loeb and Maoz 2013] in particular the James Webb Space Telescope[2] (JWST) may be able to detect biosignatures of nearby M-star planets [Seager 2015], or advanced space telescopes, such as the Darwin mission[3] and the Terrestrial Planet Finder[4]. Note however that Oxygen may be abundantly produced a-biotically [Luger and Barnes 2015]. If a few planets with *bona fide* biosignatures are found, the biotic parameter $F_b$ could be assigned an approximate value [Wandel 2015; 2016]. Quantitatively, suppose that out of a sample of $N_c$ appropriate candidates (Earth-like planets within the Habitable Zone of a star aged at least a few Gyr), biosignatures are detected in the spectra of $N_{bs}$ planets. Straightforward probability argumentation would imply that the likelihood of biotic life evolution is of the order of $F_b \sim N_{bs}/N_c$. This value should be modified to account for eventual selection effects and sampling corrections. On the other hand, if out of the above sample of $N_c$ candidate planets *no* one shows a definite biosignature, this null result would impose an upper limit $F_b < 1/N_c$. If and when the number of planets with biosignatures grows, a probability distribution function may be constructed, depending on planet lifetime, size, parent star type etc. In the near future the probability $F_b$ may be estimated observationally by spectral analyses of the atmospheres of transiting planets (e.g. Demming et al. 2009; Loeb and Maoz 2013; Wandel 2016), using telescopes such as JWST and ELT. If, following the above evidence against the Rare Earth Hypothesis, we assume $F_b$ *is not extremely small*, say, in the range of *10% < $F_b$ < 100%*, and if also M-star planets are considered suitable to host habitable planets, then $R_b \sim 10$, eq. [4a] gives $10^9$- $10^{10}$ biotic planets in our Ggalaxy, and the corresponding distance to the nearest biotic world (eq. [4c]) of the order of 10 light years.

## 3. Abundance of civilizations

In this section we extend the above analyses to putative civilizations and apply it to the Search for Extra Terrestrial Intelligence (SETI). The number of civilizations in the Galaxy ($N_c$) may be expressed by the Drake equation,

$$N_c = R^* F_s F_p F_e n_{hz} F_b F_c L_c . \qquad [5]$$

Similarly to eq. [1], the eight variables on the right hand side may be divided into two groups: the first five are astronomical factors and the last three - one biotic ($F_b$) and two "civilization" factors. The latter two are related to the development of intelligent life: the intelligence-communicative factor $F_c$, defined as the probability that one or more of the species on a planet harboring biological life will eventually develop a civilization using radio communication, and $L_c$, the broadcasting longevity (in years) of such a civilization.

Expressions analogous to eqs. [4a,b], for the number of communicative civilizations ($N_c$) and their abundance (space density, $n_c$) are easily derived by replacing $L_b$ in eqs. [3a,b] with $F_c L_c$,

$$N_c \sim 1000\, R_b F_b F_c L_{c3} \qquad [6a]$$

and

$$n_c \sim 10^{-10}\, R_b F_b F_c L_{c3}\, \text{ly}^{-3} . \qquad [6b]$$

where $L_{c3} = L_c/(1000\text{ years})$. The normalization $L_{c3}$ is just for convenience. A priori, no specific value is assumed for the longevity of radio-communicative civilizations (as is discussed below).

---

[1] http://sci.esa.int/sre-fa/47037-exoplanet-spectroscopy-mission-esm (oct 2014)/
[2] http://www.jwst.nasa.gov (oct 2014)/
[3] http://www.esa.int/Our–Activities/Space–Science/Darwin–overview oct 2014)/
[4] http://science.nasa.gov/missions/tpf (oct 2014)/

Even if communicative civilizations thrive in the Galaxy, we may be unable to detect their radio signals, unless the typical distance between them is within our detection range. As the parameter $R_b$ is relatively well estimated, and $F_b$ may be estimated in the near future, e.g. by finding biosignatures as discussed above, the major uncertainty in eqs. [6a,b] remains in the two unknown "civilization parameters", the probability for the evolution of civilizations using radio communication, $F_c$, and their longevity (the duration of their "radio loud" phase) $L_c$.

By analogy to eq. [4c] the average distance between communicative civilizations can be derived from eq. [6b],

$d_c \sim 2000 \ (R_b F_b F_c L_{c3})^{-1/3}$ ly         for $d_c < 1000$ ly         [7a]

Eq. [7a] assumes that the average distance between neighbor civilizations is smaller than the smallest dimension of the system. Since the Galaxy has a shape of a thin disk, this assumption is valid only when the distance $d_c$ is smaller than the scale height of the stellar distribution in the Galaxy (the thickness of the disk), which is ~1000 ly (in other words, as long as the product $R_b F_b F_c L_{c3} > 10$). Otherwise (if $R_b F_b F_c L_{c3} < 10$), we must take into account the flat geometry of the Galaxy. In that case we may calculate the average distance by assuming that $N_c$ civilizations are uniformly distributed on the area of the Glactic disk (a circle with a diameter of $d_G \sim 100,000$ light years ). Equating this to the area of $N_c$ circles each having a radius $d_c$ gives $\pi d_c^2 N_c \sim \pi d_G^2$ and the average distance between neighbor civilizations is $d_c \sim 10^5 N_c^{-1/2}$ ly. Substituting $N_c$ from eq. [6a] gives for this case

$d_c \sim 3000 \ ( R_b F_b F_c L_{c3} )^{-1/2}$ ly      for $d_c > 1000$ ly         [7b]

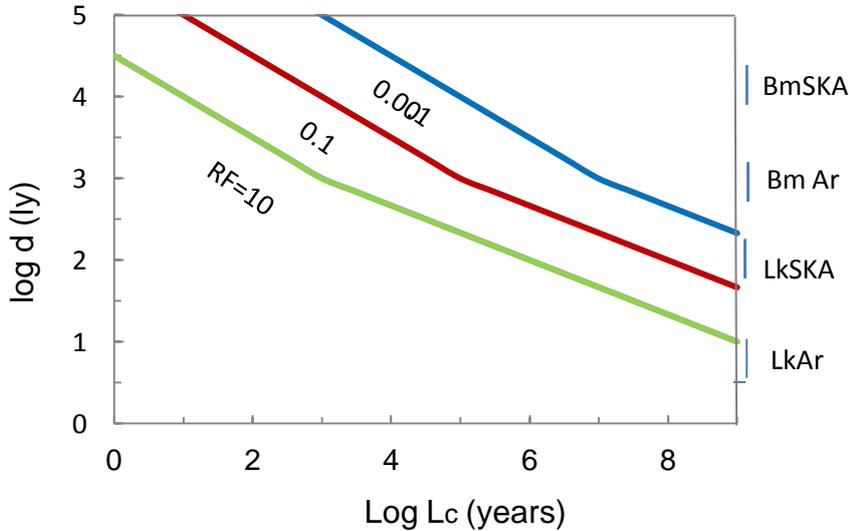

**Figure 3**. The average distance between neighbor civilizations *(d)* vs. the average longevity of a communicative civilization $L_c$, for several values of the product $RF = R_b F_b F_c$. On the vertical *right* axis are marked the relevant detection ranges of leakage and beamed radio signals by the Arecibo and SKA telescopes (see text).

Fig. 3 shows the average distance between neighbor civilizations (eqs. 7a,b) vs. the average longevity of a communicative civilization $L_c$, for several values of the product $RF = R_b F_b F_c$.

Unlike the biotic factor, the civilization parameters $F_c$ and $L_c$ cannot be inferred even intuitively from the evolution of life on Earth. As the evolution of complex life on Earth took about ~ 4 billion years, the probability for the evolution of complex and intelligent life during a few billion years could be anywhere between unity (corresponding to Earth being a typical case) and very small (if the appearance of complex life and intelligence is an extremely rare event [e.g. Cuntz et al. 2012] or typically requires a much longer time. Also $L_c$, the average longevity of a communicative civilization, cannot be inducted from its short history on Earth and could be anywhere between a few hundred years and billions of years. Communicative civilizations may disappear within a relatively short time after developing radio technology because of self destruction (wars or ecological disasters) or else become less detectable due to the development of radio-quiet communication channels. On the other hand, civilizations may survive such "childhood diseases" by spreading to other planets and last much longer. The weighted average of the longevity could be increased by such older, long lasting civilizations which broadcast for long time.

**4. 1 The SETI probability**

There are two scenarios for detecting radio signals from extraterrestrial civilizations: (1) finding a purposeful, directed broadcast attempt, including an interstellar automatic radio beacon or (2) civilizations may be detected through no special efforts of their own. The latter hypothesis, often called eavesdropping, is concerned with the extent to which a civilization can be unknowingly detected through the by-products of its daily activities, e.g. the leakage of its own radio communication to space. The range for detecting such radio signals depends on the receiver sensitivity and on the transmitting power, as well as on the level of the background noise and on whether the signal is directed or isotropic. Beamed transmissions directed at a specific target would be much stronger and thus detectable from longer distances than the semi-isotropic broadcasting, typical for radio stations, looked after by eavesdropping. Future radio observatories such as EVLA (Expanded Very Large Array), LOFAR (Low Frequency Array) and Square Kilometer Array (SKA) may be able to detect low-frequency radio broadcast leakage from a civilization with a radio power similar to ours out to a distance of a few hundred of light years [Loeb and Zaldarriaga, 2007]. SKA would be able to detect an airport tower radar from 30 light years[5]. Beamed transmissions could be detected over much larger distances. For example, a targeted search by the Arecibo telescope could detect alien signals sent by a similar device (i.e., with a similar power, $\sim 10^{13}$ Watt/m$^2$/radian$^2$) and aimed at Earth from distances of a few thousands of light years. Noteworthy, the sensitivity of all-sky surveys is much lower and the above detection ranges need to be decreased by a factor of 10 -- 100. Eqs. [7a,b] imply that even with rather "optimistic" values of the other parameters ($R_b$ $F_b$ $F_c \sim 1$), unless $L_c$ is longer than a million years, the average distance between neighbor civilizations is thousands of light years, far beyond the range of eavesdropping even by future telescopes such as SKA.

Transmissions beamed at Earth, either unintentionally, like an aviation radar or communication satellite, or intentionally like the Arecibo message of 1974[6], may be detected from considerably larger distances. For example, a beamed transmission at the broadcasting power similar to that of the Arecibo radar can be detected by the Arecibo radio telescope at a range of $\sim$ 3000 light years, and by future telescopes such as SKA the detection range may increase to $\sim$30,000 light years, virtually across the Galaxy. These ranges are shown on the right vertical axis in fig. 3. Note however that the effective broadcasting time of beamed signals may be significantly shorter than the total communicative longevity, as discussed below.

The expressions for the average distance between civilizations derived above can be used to estimate the success chances of SETI by calculating the probability that a broadcasting civilization happens to lie within the detection range of present and future radio telescopes. Let us first consider eavesdropping (looking for leakage signals). Using the sensitivity of future radio telescopes such as SKA, the eavesdropping detection range is of the order of 100 light years. Equation [7a] shows that a neighbor civilization is likely to exist within this distance if the "Drake product" $F_b$ $F_c$ $L_c$ is of the order of a million years or more. For smaller values of the Drake product we may define the likelihood $p(d)$, that a civilization happens to exist at a distance $d < d_c$, that is, closer to Earth than the average distance between neighboring civilizations. A straightforward geometric approach (fractional volume) gives

$$p(d) \sim (d/d_c)^3. \qquad [8]$$

---

[5] /https://www.skatelescope.org/key-documents (oct. 2014)
[6] http://www.seti.org/seti-institute/project/details/arecibo-message (nov. 2014)

Fig. 4 shows the likelihood $p(d<100)$, that a broadcasting civilization happens to exist within a distance of 100 ly from Earth, as a function of the average longevity of communicative civilizations, for several values of the Drake product.

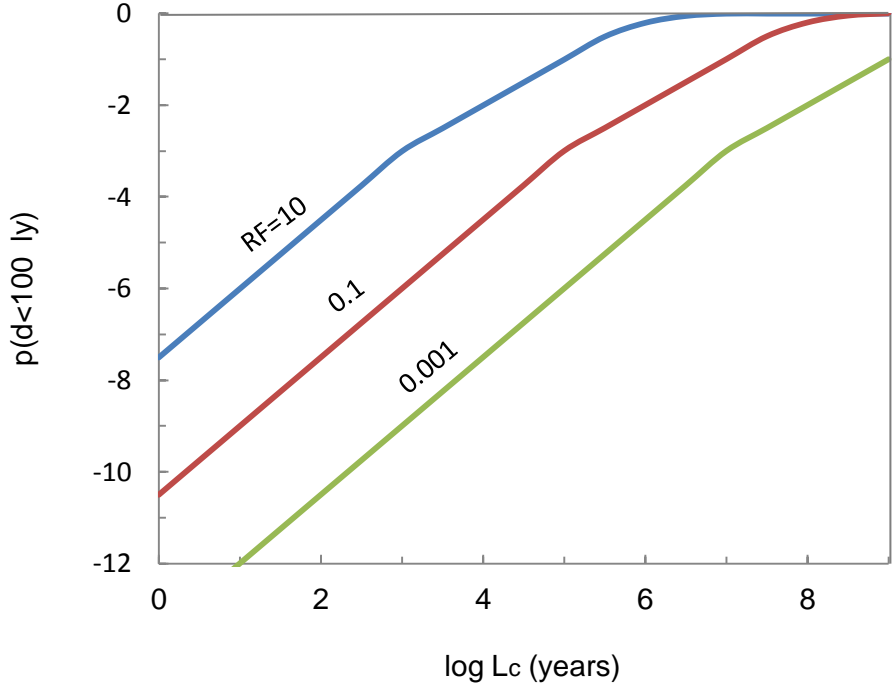

**Figure 4**. The probability of a civilization to exist within a distance of 100 light years from Earth, vs. the average longevity of a communicative civilization, $L_c$, for several values of the product $RF = R_b F_b F_c$.

Combining eqs. [8] and [7a,b] gives the probability to detect leakage signals in terms of the Drake parameters (assuming a detection range of 100 ly, appropriate for planned radio arrays such as SKA):

$$p(d<100\text{ ly}) \sim 10^{-4} R_b F_b F_c L_{c3} \qquad d_c < 1000\text{ ly}, \qquad [9a]$$

$$p(d<100\text{ ly}) \sim 3\,10^{-5} (R_b F_b F_c L_{c3})^{3/2} \qquad d_c > 1000\text{ ly}. \qquad [9b]$$

As in eq. [7a], the condition on $d_c$ in eq. [9a] is equivalent to a condition on the product $R_b F_b F_c L_{c3} > 10$, and in eq. [9b] $R_b F_b F_c L_{c3} < 10$. For example, if we assume that communicative civilizations are common ($R_b F_b F_c \sim 1$), and if they typically transmit for ten thousand years ($L_{c3} \sim 10$), the probability that a civilization is presently broadcasting within ~ 100 ly from Earth is 0.001.

**4.2 The probability of detecting beamed signals**

As discussed above, assuming the broadcasting power of putative civilizations is comparable to ours, eavesdropping on leakage signals has a quite limited range; the Arecibo telescope could detect leakage signals from civilizations broadcasting at the power of Earth at a distance of only a few light years, and future telescopes such as SKA up to ~100 ly. On the other hand, beamed transmissions may be detected from considerably larger distances. For example, a beamed transmission comparable to that of the Arecibo radar may be detected by the Arecibo dish at a range of ~ 3000 ly, and by future telescopes such as SKA up to ~30,000 ly. Apparently, these increased detection ranges should give a significantly higher detection probability. On the other hand, beamed signals may not be continuously aimed in our direction, contrary to leakage signals. This effect is likely to reduce the effective broadcasting duration. It may be described by introducing a "beaming parameter" $b$, the fraction of the broadcasting lifetime of a civilization during which it is actually sending signals beamed in our direction. This beaming may happen either unintentionally, as communication with satellites, spacecrafts, or planets in their own solar system, or deliberately as interstellar messages [e.g. Zaitsev 2006]. In other words, $bL_c$ is the integrated duration of beamed broadcasting in the direction of Earth.

For $R_b F_b F_c b L_{c3} < 10$ (see eq. 9b) the probability to find a civilization within the detection range of beamed signals by Arecibo end SKA, respectively, is

$p(d< 3000 \text{ ly}) \sim 1 \, ( R_b F_b F_c b L_{c3} )^{3/2}$      Arecibo,      [10a]

and

$p(d<30,000 \text{ ly}) \sim 1000 \, ( R_b F_b F_c b L_{c3} )^{3/2}$      SKA.      [10b]

For example, assuming that communicative civilizations are common, that is $R_b F_b F_c \sim 1$, and that on average signals are beamed at Earth during an integrated time of $bL_c \sim 10$ years, equations [10a,b] give a detection probability $p \sim 10^{-3}$ by Arecibo and $p\sim 1$ by SKA. If actually $R_b F_b F_c b L_{c3} <10$ it is not surprising that SETI has not yet detected an alien signal (the "Great Silence"), and it may remain silent even with the increased sensitivity of future telescope arrays. If, on the other hand, $R_b F_b F_c b L_{c3} >10$, then applying eq. [9a] to beamed signals gives

$p(d< 3000 \text{ ly}) \sim 10^{-4} R_b F_b F_c b L_{c3}$      Arecibo,

and

$p(d<30,000 \text{ ly}) \sim 1$      SKA,

implying that SKA might be able to detect signals beamed at Earth by putative civilizations.

**4.3 Estimating the Civilization Parameters**

Similarly to the biotic factor, also the civilization parameters $F_c$ and $L_c$ are presently unknown. If SKA and advanced future radio arrays fail to detect intelligent extraterrestrial signals, eqs. [9a,b] and [10b] may be used to place an upper limit on the products $F_c L_c$ and $F_c bL_c$, respectively. On the other hand, if an extraterrestrial signal from another civilization is detected, the present ignorance in the civilization parameters may be removed, or at least significantly constrained. If by then also biomarkers of exoplanets are detected, so that $F_b$ can be estimated to some extent, the product $F_c L_c$ could be assigned an approximate value; suppose that out of a sample of $N_{bs}$ suitable civilization candidates (that is, Earth-like planets aged at least a few Gyr with a biosignature) intelligent signals are detected from $N_{is}$ ones; here, in analogy to the section on the biotic parameter above, $N_{bs}$ is the number of planets with biotic signature, and $N_{is}$ is the number of planets from which intelligent signals are detected. A similar argument as in the case of the biotic factor would count for the "communicative factor", or the likelihood of a communicative civilization to evolve on a biotic planet, multiplied by the average broadcasting longevity, leading to a value of the order of

$F_c L_c \sim 10 \text{ Gyr } N_{is}/N_{bs}$,

where 10 Gyr is the age of the Galaxy. As in the case of the biotic fator, this expression needs to be modified by eventual selection effects and sampling corrections, as well as by putative contributions to $N_{is}$ from non biotic planets such as automated beacons.

5. Summary

The recent results of the Kepler mission significantly reduce the uncertainty in the astronomical parameters of the Drake equation. I derive expressions for the space density of biotic worlds as a function of the (yet unknown) probability for the evolution of biotic life and the uncertainty in the astronomical parameters. Similar expressions are derived for the distance to putative communicative civilizations, depending on two additional unknown factors in the Drake equation, the probability of evolution from simple biotic life to a communicative civilization and its longevity. Additionally, the probability to detect radio signals from other civilizations with present and future radio telescopes is estimated in terms of these factors. The extended analyses, updated by the Kepler results, presented in this paper suggests that our nearest biotic neighbor exoplanets may be as close as 10 light years. Even with a less optimistic estimate of the biotic probability, for example that biotic life evolves on one in a thousand suitable planets, our biotic neighbor planets may be expected within 100 light years. On the other hand, the distance to the nearest putative civilizations, even for optimistic values of the Drake parameters, is estimated to be thousands of light years.